\documentclass[aps,pra,a4paper,twocolumn,superscriptaddress,epsf,showpacs,showemail]{revtex4}

\usepackage{graphicx}
\usepackage{dcolumn}
\usepackage{bm}
\usepackage{amsmath}

\begin{document}


\title{Dynamically Controlled Toroidal and Ring-Shaped Magnetic Traps}



\author{T. Fernholz}
\email{tfernhol@science.uva.nl}
\author{R. Gerritsma}
\affiliation{Van der Waals-Zeeman Institute, University of
Amsterdam, 1018 XE Amsterdam, The Netherlands}
\author{P. Kr\"uger}
\affiliation{Laboratoire Kastler-Brossel, CNRS, Ecole Normale
Sup\'erieure, 75005 Paris, France} \affiliation{Physikalisches
Institut, Universit\"at Heidelberg, 69120 Heidelberg, Germany}
\author{ R. J. C. Spreeuw}
\affiliation{Van der Waals-Zeeman Institute, University of
Amsterdam, 1018 XE Amsterdam, The Netherlands}


\date{\today}

\begin{abstract}

We present traps with toroidal $(T^{2})$ and ring-shaped
topologies, based on adiabatic potentials for radio-frequency
dressed Zeeman states in a ring-shaped magnetic quadrupole field.
Simple adjustment of the radio-frequency fields provides versatile
possibilities for dynamical parameter tuning, topology change, and
controlled potential perturbation. We show how to induce toroidal
and poloidal rotations, and demonstrate the feasibility of
preparing degenerate quantum gases with reduced dimensionality and
periodic boundary conditions. The great level of dynamical and
even state dependent control is useful for atom interferometry.

\end{abstract}



\maketitle



\section{Introduction}


Cold neutral atoms are particularly well suited as a generic model
for many-body systems as they allow tuning of parameters that are
not accessible in other systems such as the interparticle
interaction strength. 
Studying the properties of quantum gases in confining geometries
where the dimensionality of the system can be chosen or even be
adjusted is at the focus of current research \cite{esslinger}. The
dimensionality of the confined cloud plays a key role for
important properties such as long range order \cite{mermin,
bagnato} and superfluidity
\cite{berezinskii,kosterlitz,castin,hadzibabic} of the gas. This
is also true for the topology of the trap, and hence large
interest in multiply connected trapping potentials such as rings
has recently emerged.

In first experiments with nontrivial trapping topologies,
ultracold and Bose-condensed atoms have been confined in
ring-shaped potentials \cite{chapman,stamperkurn,riis}. Most of
these experiments rely on the production of a ring-shaped magnetic
quadrupole field to trap the atoms. The ring of zero field in the
center of this trap causes a loss of atoms due to Majorana spin
flips to untrapped states. These leaks have been plugged using
time orbiting ring traps (TORTs) \cite{stamperkurn}, out-of-plane
current carrying wires \cite{riis}, or stadium-shaped traps
reducing the ring of zero to four points \cite{prentiss}. Also
electrostatic fields have been proposed as a possible solution for
a chip-based trap \cite{hopkins}. For chip-based approaches
\cite{hopkins, crookston, vengalattore} the symmetry-breaking
potential corrugation caused by lead wires must typically also be
addressed.

In this paper, we show how adiabatic potentials for
radio-frequency- (rf-) dressed states \cite{garraway, perrin,
schumm, lesanovsky, courteille, white, morizot} can be used in a
circular configuration. These dressed traps have recently been
shown to provide long lifetimes, allow for evaporative cooling
\cite{alzar, hofferberth}, and have been used to coherently split
matter waves \cite{schumm, ketterle}.

We present an experimentally feasible method that results in a
trap with new flexibility going well beyond that of a TORT trap,
using a similar setup but with an oscillating field of much lower
amplitude at higher frequencies. Reaching beyond simple ring
geometries, toroidal surfaces can be formed and smoothly
transformed to a (multiple) ring geometry. One- and
two-dimensional quantum gases with periodic boundary conditions
are accessible as well as the crossover regimes between
confinement of different dimensionality (3D to 2D and 1D, and 2D
to 1D). Moreover, we show that the trapped gases can be
dynamically set into various types of motion by simple
experimental techniques. These include toroidal and poloidal
rotations on torus surfaces and independently controlled
acceleration along circular paths.

\section{Trap design}


\begin{figure}
\includegraphics[width=3.4 in]{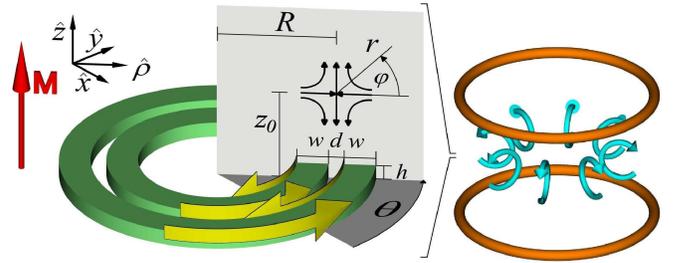}\\
\caption{(Color online) Schematic view of a quadrupole ring
produced by two rectangular rings of uniform magnetization
$M\cdot\mathbf{\hat{z}}$, mean radius $R$, widths $w$, distance
$d$, and height $h$. Additional coordinates are given by the
distance $r$ from the ring of zero field, and the toroidal and
poloidal angles $\theta$ and $\varphi$, respectively. An
additional pair of external coils is used to produce a circularly
polarized, axially symmetric rf field. \label{figRingDim}}
\end{figure}

Our trapping configuration is based on magnetic micro-structures,
which have been proposed and demonstrated for atom trapping both
using current-carrying wires \cite{schmiedmayer1, reichel1,
dekker} and permanent magnetic structures \cite{hinds1,
spreeuw1,sidorov1}. We start from a ring-shaped magnetic
quadrupole field, generated by two concentric rings of magnetized
material with out-of-plane magnetization $\mathbf{M}$; see
Fig.~\ref{figRingDim}. The advantages here are high field
gradients (up to $10^{3}$ T/m) and the absence of lead wires that
break the symmetry of the trap. For small heights $h$ of the
magnetic layer, the field sources become equivalent to two pairs
of concentric, counterpropagating, linelike currents $I=M h$
around the edges of the material. They produce a magnetic
quadrupole field with a ring-shaped line of zero field. We assume
that the radii of the rings are large compared to their widths and
separation, $R\gg w,d$. In this case, the field in the vicinity of
the magnetic rings can be approximated by a two-dimensional
quadrupole field with field gradient $q$, centered above the mean
radius $R$ at height $z_{0}=\frac{1}{2}\sqrt{d(2w+d)}$. Its axes
are parallel and perpendicular to the surface.


Cold atoms in a low-field-seeking spin state can be confined in
this field, but will suffer losses due to nonadiabatic spin-flip
transitions near the line of zero field. These can be avoided
using adiabatic, radio-frequency-induced potentials. A theoretical
description was developed in \cite{garraway,perrin}. Resonant
coupling between Zeeman levels leads to a modification of the
dressed state energies, producing local potential minima near
positions where the coupling field is resonant with the atomic
Larmor frequency. As was noted in \cite{schumm} and recently
discussed in \cite{lesanovsky} for a cylindrical case, the
polarization of the fields must be taken into account. This
important design parameter introduces new possibilities, on which
we put emphasis in this paper.

For an atom with total angular momentum $F$, the Hamiltonian in a
weak magnetic field can be approximated by
$H(t)=g_F\,\mu_B\,\hat{\mathbf{F}}\cdot\mathbf{B}(t)$, with the
Bohr magneton $\mu_B$ and the Land\'e factor $g_F$. The magnetic
field consists of a static contribution $\mathbf{B}_\mathrm{dc}$
as given by the ring-shaped quadrupole field (Fig.
\ref{figRingDim}) plus an rf contribution
$\mathbf{B}_\mathrm{rf}(t)=\textrm{Re}[\mathbf{B}_\mathrm{rf}\exp(-
i\omega t)]$. At each point we take the direction of the static
field as the local quantization axis,
$\mathbf{B}_\mathrm{dc}=B_0\mathbf{\hat{e}}_0$.

The vector $\mathbf{B}_\mathrm{rf}$ is complex valued and can be
decomposed in a basis of orthonormal, spherical polarization
vectors $(\mathbf{\hat{e}}_0, \mathbf{\hat{e}}_+,
\mathbf{\hat{e}}_-)$, corresponding to $\pi$, $\sigma^+$, and
$\sigma^-$ polarizations, respectively: $\mathbf{B}_\mathrm{rf}=
\beta_0\mathbf{\hat{e}}_0+\beta_+\mathbf{\hat{e}}_++\beta_-
\mathbf{\hat{e}}_-$. We transform the Hamiltonian to a frame
rotating about $\mathbf{\hat{e}}_0$ at the rf frequency $\omega$
and make a rotating-wave approximation (RWA) by neglecting
remaining time-dependent terms. The resulting effective
Hamiltonian in the rotating frame has the eigenvalues:
\begin{equation}
E_{m_F}=m_F\, \mu_B\,g_F
\sqrt{(B_0-\hbar\,\omega/g_F\,\mu_B)^2+|\beta_+|^2/2}
\label{eqEigenvalues}.
\end{equation}
It is crucial that in the RWA only the $\sigma^{+}$ field
component (relative to the static field) appears in the effective
Hamiltonian.



\begin{figure}
\includegraphics[width=3 in]{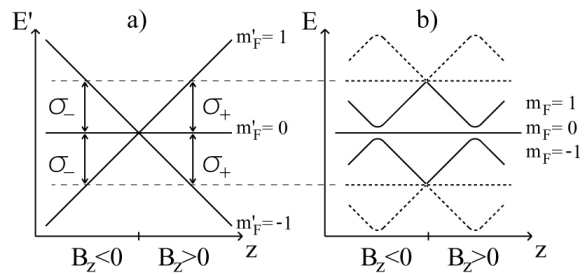}\\
\caption{Schematic view of rf-induced potentials: (a) The energy
splitting between atomic Zeeman levels depends linearly on
position in a magnetic quadrupole field. (b) The energy levels
appear shifted in the rotating frame and exhibit avoided crossings
at resonant rf coupling. Note that across $B_{z}=0$ the effective
quantum number $m_{F}$ changes sign and is undefined at the field
zero. Atoms traversing this region can undergo transitions to
different spin states in the adjacent dressed-state manifolds
(dashed lines). For negative $B_{z}$ or negative $g_F$ the
resonance frequency is also negative, inverting the rotational
sense of the required coupling field when referenced to the
z-direction. \label{figCoupling}}
\end{figure}

For a ring-shaped quadrupole field, the resonance condition is met
on a toroidal surface of constant field modulus
$B_0=q\,r_0=\hbar\,\omega/g_F\,\mu_B$. In Fig.~\ref{figCoupling}
the trapping potential along the $z$-direction at $\rho=R$ is
depicted. Atoms can be trapped at the avoided crossings between
dressed energy levels, which occur on the surface of resonance.
The remaining potential on this surface is determined by the rf
component $\beta_+$, referenced to the local direction of the
static field.


In order to form a trap, $\beta_+$ must not vanish in any point on
the resonant surface. Such points, like the zeroes in quadrupole
magnetic traps, limit the trap lifetime and have to be avoided. It
is not sufficient to use any rf field that provides field
components perpendicular to the static field, as, e.g., a rotating
field in the $\mathbf{\hat{x}}, \mathbf{\hat{y}}$-plane fails. A
good way to drive the resonance is by a phase-shifted
superposition of a spherical quadrupole field, radially polarized
along $\boldsymbol{\hat{\rho}}$ and a uniform field with linear
$\mathbf{\hat{z}}$ polarization. A single pair of external coils
can generate this field without breaking the axial symmetry. The
coils can be placed symmetrically above and below the chip
surface. An interesting possibility to generate phase-shifted
radial fields is to incorporate conducting metal rings below the
surface and induce rf currents. In general, the rf field is
elliptically polarized in the $\boldsymbol{\hat{\rho}},
\mathbf{\hat{z}}$-planes. Denoting its circular components by
complex amplitudes $a$ and $b$, it is expressed in global,
Cartesian coordinates by:
\begin{equation}
\mathbf{B}_{1}= \frac{a}{\sqrt{2}} \left(\begin{array}{c}
\cos\theta\\
\sin\theta\\
i\\
\end{array}\right)+
\frac{b}{\sqrt{2}} \left(\begin{array}{c}
\cos\theta\\
\sin\theta\\
-i\\
\end{array}\right)
\end{equation}

In the vicinity of the quadrupole ring the static field direction
is $\mathbf{\hat{e}}_0=(-\cos\theta\,\cos\varphi,\,
-\sin\theta\,\cos\varphi,\, \sin\varphi)$. We construct a
right-handed orthonormal triplet by choosing the direction
tangential to the ring, $\mathbf{\hat{e}}_1=(-\sin\theta,
\cos\theta, 0)$, and
$\mathbf{\hat{e}}_2=\mathbf{\hat{e}}_0\times\mathbf{\hat{e}}_1$.
We define:
\begin{equation}
\mathbf{\hat{e}}_+=\frac{(\mathbf{\hat{e}}_1+i\,\mathbf{\hat{e}}_2)}{\sqrt{2}}=
\frac{1}{\sqrt{2}}\left(
\begin{array}{c}
-\sin\theta -i\cos\theta\sin\varphi \\
\cos\theta -i\sin\theta\sin\varphi \\
-i\cos\varphi
\end{array}
\right), \label{eqSigmaComponent}
\end{equation}
and obtain the coupling component:
\begin{equation}
\beta_+=\mathbf{\hat{e}}_+^*\cdot \mathbf{B}_1=-\frac{a}{2}\,
e^{-i\varphi} +\frac{b}{2}\, e^{i\varphi}.
\end{equation}
The potential on the surface of resonance is given by:
\begin{equation}
E_{m_F}(\varphi)=\frac{\mu}{2\sqrt{2}}
\sqrt{|a|^{2}+|b|^{2}-2|a||b|\cos\left[2(\varphi-\varphi_0)\right]},
\label{eqPoloidalPotential}
\end{equation}
where $2\varphi_0=\arg b-\arg a$ and $\mu=m_F\,g_F\,\mu_B$.


\begin{figure}
\includegraphics[width=3.1 in]{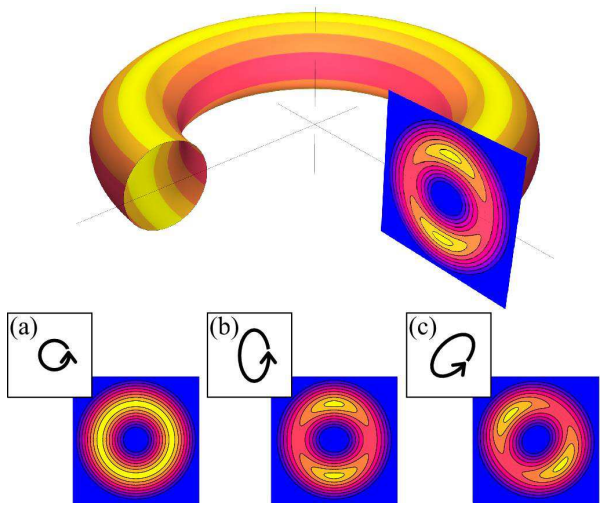}\\
\caption{(Color online) Visualization of the adiabatic trapping
potential. On the top, the potential on the resonant torus for
vertically split rings is shown together with a cut parallel to a
$\rho,z$-plane. On the bottom, $\rho,z$-potentials are shown for
(a) circular rf ($b=0$), and elliptical rf ($a>b$) with (b)
$\varphi_{0}=\pi/2$, and (c) $\varphi_{0}=\pi/4$ (increasing
potential from light to dark).\label{figTorusPotential}}
\end{figure}

We assume $R\gg r_{0}$ and thus approximate $a$ and $b$ to be
independent of $r$ and $\varphi$. A special case is obtained for a
circular field, $a=0$ or $b=0$. The potential,
Eq.~(\ref{eqPoloidalPotential}), then becomes independent of both
$\varphi$ and $\theta$, so the trap extends over the entire
toroidal surface; see Fig.~\ref{figTorusPotential}a). This opens
up the possibility to trap cold atoms in a two-dimensional,
boundaryless geometry.

More generally, if neither $a$ nor $b$ vanishes, the field is
elliptically polarized. In this case, the potential has minima for
two poloidal angles. The trap thus splits poloidally into two
opposite rings, as shown in Fig.~\ref{figTorusPotential}. The
orientation of this splitting is determined by $\varphi_0$.

For convenience, we redefine the local coupling amplitude in the
poloidal minimum as $c=\big||a|-|b|\big|/2\sqrt{2}$ and define
$\epsilon=|b/a|$. The trapping frequencies in the radial and
poloidal directions then become:
\begin{equation}
\omega_r=\frac{q}{\sqrt{c}}\;\sqrt{\frac{\mu}{m}}, \qquad
\omega_{\varphi}=
\frac{2}{r_0}\;\sqrt{\frac{c\,\epsilon}{(1-\epsilon)^2}}\;\sqrt{\frac{\mu}{m}},
\label{eqTrapfrequencies}
\end{equation}
where $m$ is the mass of an atom.

\section{Adiabaticity requirements}

The condition that the internal atomic state adiabatically follow
the external motion is violated at points where the rf field is
resonant but the coupling component $\beta_+$ is too small. For
example, for $|a|=|b|$ - i.e., linear rf polarization - the
coupling will vanish in the two rings. To avoid Landau-Zener
transitions, the energy splitting $E_{m_F}/m_F$ must be large
compared to the trap frequency $\omega_r$. We thus demand that
$E_{m_F}/m_F\geq\alpha\,\hbar\,\omega_r$, with the factor $\alpha
\gg 1$ and get the adiabaticity condition:
\begin{equation}
\frac{c^3}{q^2}\geq\frac{\alpha^2\,\hbar^2\,m_F }{\mu_B\,g_F\,m}.
\label{eqAdiabatic}
\end{equation}

In addition, it is necessary to stay in the RWA validity regime;
i.e., the rf frequency must be high compared to the energy
splitting. Thus we choose $\hbar\,\omega\geq\delta\,E_{m_F}/m_F$
with the numerical factor $\delta > 1$. This is equivalent to
$q\,r_0\geq\delta\,c$. Both relations together impose a size
restriction \footnote{This relation separates the setup from a
TORT; a continuous transition between the trap types is
impossible.}:
\begin{equation}
q\,r_0^3\geq\frac{\alpha^2\,\delta^3\,\hbar^2\,m_F}{\mu_B\,g_F\,m}.
\label{eqDeep}
\end{equation}

\section{Low-dimensional traps}

As a first application we discuss the feasibility to realize
two-dimensional traps for degenerate quantum gases. One
requirement to be met is that the thermal motion of the atoms be
limited to the radial ground state,
$\hbar\,\omega_r\geq\zeta\,k_B\,T$ with $\zeta>1$. Together with
the adiabaticity condition, the achievable temperature sets lower
limits to the fields:
\begin{equation}
q\geq\sqrt{\frac{\alpha\,m}{m_F}}\cdot\frac{(\zeta\,k_B\,T)^{3/2}}{g_F\,\mu_B\,\hbar},
\qquad c\geq\frac{\alpha\,\zeta\,k_B\,T}{g_F\,\mu_B}.
\end{equation}
With $\zeta\,T=1~\mu$K, $\alpha=10$, and $^{87}$Rb atoms in the
$|F,m_F\rangle=|2,2\rangle$ state, the required fields are
$q>90$~T~m$^{-1}$, which is easily achievable with
micro-structured permanent magnetic material, and $c>30~\mu$T.
Note that $c$ plays a similar role as the minimum field in a
Ioffe-Pritchard magnetic trap, typically $\sim 100~\mu$T.
Neglecting angular momenta, we estimate the lifetime in the trap
by a simple integration of the Landau-Zener spin-flip probability
over the radial velocity distribution. For the given numbers we
find a negligible loss rate. Under our assumptions and contrary to
the behavior in more common traps with confinement along two or
three dimensions, the lifetime should strongly decrease for higher
temperatures as all atomic trajectories necessarily cross the trap
minimum twice per radial oscillation period. The estimated
lifetime reaches 1~s for $T\approx 100~\mu$K, allowing for a final
evaporative cooling stage in this trap.

The second requirement is that atomic collisions do not excite
radial motion. This is ensured when the chemical potential
$g_{2d}\,n_{2d}\approx
(2\,\hbar)^{3/2}\,(\pi\,\omega_{r}/m)^{1/2}\,a_s\,N/A$ is lower
than $\hbar\,\omega_{r}$, where $a_s$ is the s-wave scattering
length.

A unique feature of two dimensional systems is the occurrence of
superfluidity in a cold temperature phase without spontaneous
symmetry breaking. The underlying microscopic mechanism of the
Berezinskii-Kosterlitz-Thouless type is currently under
investigation in harmonically trapped cold atomic clouds
\cite{hadzibabic}. A toroidal trapping geometry as discussed here
allows to approximate the homogeneous case more closely. While a
true Bose-Einstein condensate in two dimensions is still possible
in harmonically confined two-dimensional gases, this is no longer
true if the potential is homogeneous in both dimensions
\cite{bagnato}. In an infinite system, the
Berezinskii-Kosterlitz-Thouless transition temperature is
proportional to the atomic density:
$T_{BKT}=(\pi\hbar^{2}/2m\,k_{B})\,N/A$
\cite{berezinskii,kosterlitz}. Here we equate the superfluid
density $\rho_s$ and the total density $\rho_t$. The difference
between $\rho_s$ and $\rho_t$ and recent measurements on the
critical temperature in a harmonically trapped system are
discussed in Ref. \cite{kruger}. To achieve a high enough
transition temperature at reasonable atom numbers, the toroidal
surface $A\approx 4\pi^2\,R\,r_0$ must be chosen as small as
possible. The minimal radius $r_{0}$ depends on the static field
gradient $q$ and is given by Eq.(\ref{eqDeep}). The radius $R$
cannot be chosen arbitrarily small as the circular rf amplitude
$a$ is limited by its radial quadrupole component. We find that it
is favorable to use lower static field gradients $q$. This
requires larger $r_{0}$ ($\sim q^{-1/3}$), but allows for smaller
$R$ as it reduces the required rf amplitude $a$ ($\sim q^{2/3}$).
The limit to this strategy is given by the decreasing trap
frequency $\omega_{r}$. Defining the radial rf gradient
$p=|a|/(\sqrt{2}\,R)$ and requiring
$\hbar\,\omega_{r}\geq\zeta\,k_{B}\,T_{BKT}$, we find the optimum
static field gradient:
\begin{equation}
q=\frac{\zeta\,N\,p}{16\,\pi\,\alpha\,\delta\,m_{F}}.
\end{equation}

With $N=10^5$ $^{87}$Rb atoms in the $|F,m_F\rangle=|2,2\rangle$
state, $\alpha=10$, $\delta=\zeta=5$, and $p=1$~T~m$^{-1}$, a
transition temperature of $T_{BKT}\approx 215$~nK can be achieved
at torus sizes of $R\approx 65~\mu$m and $r_0\approx 1.6~\mu$m
with $q\approx 100$~T/m, and $\omega=2\pi \times 1.1$~MHz. With
these parameters, we estimate the chemical potential to be well
below the trap frequency, $g_{2d}\,N/A<\hbar\,\omega_{r}$. For an
accurate description, the finite size of the system must be taken
into account. But since the poloidal circumference is still more
than 40 times longer than the healing length of the gas
$\xi=\hbar/(2\,m\,g_{2d}\,n_{2d})^{1/2}\approx 225$~nm, we expect
the above parameters to be a good approximation for the crossover
to superfluidity.

For this kind of geometry to be implemented it is essential to
compensate gravity over the vertical extension of the torus, $2
r_0$. The resulting potential variation over this scale must be
small compared to the chemical potential. In the above example the
gravitational potential difference amounts to $\Delta
U_{gr}\approx h\times 7$~kHz, while the chemical potential is only
$g_{2D}n_{2D}\approx h\times 1$~kHz. This problem can be solved in
various ways.

It is straightforward to show that an rf field parallel to
$\mathbf{\hat{e}}_1$, as produced by a straight central wire
perpendicular to the plane, could be used to cancel the vertical
potential difference. To implement such a solution, however, is
technologically demanding. A simplified approach could be based on
a short-circuited coaxial cable mounted vertically below the chip
surface. A second solution is the use of weak electro-static
fields. We estimate that applying voltages on the order of
$\approx 1$~V to a micrometer-sized structure at a distance of
$\approx 10~\mu$m creates potential gradients comparable to
gravitation \cite{krueger}. Another solution is to reflect a
far-detuned laser beam under grazing incidence off the metallic
chip surface. The optical dipole force at the height of the torus
compensates gravity already at moderate laser intensity. The
grazing incidence is required for maximal linearity, because it
creates a long wavelength standing-wave pattern above the chip
surface. By varying the incident angle, the optical lattice period
can be chosen such that the torus's height $z_0$ is at the first
or second inflection point of the dipole potential. A feasible
example is to reflect a red-detuned laser ($\lambda = 1064$~nm) at
a grazing angle of $40$~mrad off the chip surface ($13.3~\mu$m
lattice period, $z_0=10~\mu$m). An intensity of
$I=920$~mW/mm$^{2}$ suffices to compensate gravity with a
negligible photon scattering rate ($\approx 10^{-3}$~s$^{-1}$).
The remaining nonlinear vertical potential variation amounts to
$\Delta U\approx h\times 200$~Hz. Focusing the laser onto a size
somewhat larger than the torus [full width at half maximum (FWHM)
$w_0\sim 700~\mu$m] ensures homogeneous illumination ($\Delta
U_{dipole}<h\times 200$~Hz) and requires a laser power of $P\sim
500$~mW, which can be reduced using astigmatic focussing. Note
that the given example is adapted to a trap below the chip.
Compensation of gravity for a trap above the chip surface requires
a blue-detuned laser or a smaller incidence angle to reach the
first inflection point of the dipole potential.

It is possible to reduce the dimensionality further and create
one-dimensional rings. In this case also the poloidal degree of
freedom must be frozen, $\hbar\,\omega_{\varphi}\gg
k_{B}\,T,g_{1d}\,n_{1d}$. At reasonable rf amplitudes of
$a,b\approx 10^{-4}$~T, this can be achieved with the radius
$r_{0}$ chosen in the micron range. It enables studies of
superfluidity in one-dimensional gases \cite{castin}.

In macroscopic ring geometries it has not yet been possible to
create a complete ring-shaped condensate as potential corrugations
were larger than the chemical potential of the atomic cloud
\cite{stamperkurn}. Recently, the corrugations near microscopic
permanent magnetic trapping structures have been studied in
detail, showing similar magnitudes \cite{whitlock}. For
microscopic traps progress in the fabrication process might be
expected and the problem is less severe because larger chemical
potentials can be reached with tighter confinement. In addition,
the rf dressing of the potential reduces the corrugations in the
effective potential experienced by the atoms, because atoms are
trapped at positions of constant field modulus. A perturbing field
therefore displaces the trap, but the potential minimum is
determined by the coupling to the rf field. The coupling is
altered by a tilt of the static field with respect to the
polarization of the rf field. In our case, only the corrugating
field component tangential to the toroidal direction $\Delta
B_{\theta}$ modifies the potential minimum. We find that for small
perturbations the potential corrugation is given by $\Delta
E\approx \mu c (\frac{\Delta B_{\theta}}{q r_0})=\mu \Delta
B_{\theta}/\delta$. Compared to the undressed case, the effect of
corrugations is suppressed by a factor of $\delta$.

\section{Controlling atomic motion}

\begin{figure}
\includegraphics[width=3.2 in]{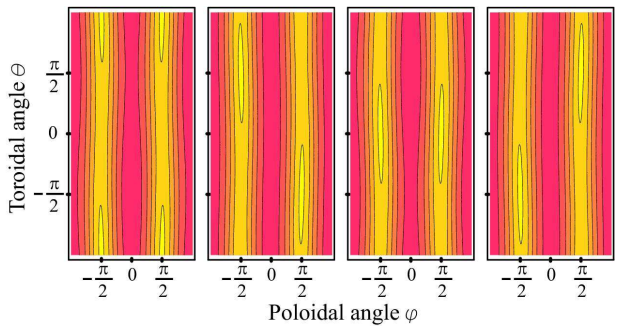}\\
\caption{(Color online) Contour plots of the trapping potential in
unfolded $\theta,\varphi$-coordinates on the resonant torus for
vertically split, counter-rotating rings. Potentials are shown for
$\arg u=\arg v=0$, $\pi/2$, $\pi$, and $3\pi/2$ (increasing
potential from light to dark).\label{figPotCounterRings}}
\end{figure}

As a second application we consider the possibilities for
controlling atomic motion by slowly varying and weakly perturbing
the rf fields. As the relative phase of $a$ and $b$ controls the
poloidal angle $\varphi_0$ of the trap minima, a small frequency
difference between the two components can be used to induce
rotation of atoms around the quadrupole center. Such stirring in
the poloidal direction can be used to excite vortex loops that lie
inside the toroidal surface.

It is also possible to produce and rotate trap minima in the
toroidal direction by additional fields. Split rings can even be
controlled independently from each other. As an example, we
discuss counterrotating, vertically split rings. After inducing
such a splitting with $\mathbf{B}_1$, the rotations can be
introduced using a weaker, linearly polarized rf field, parallel
to the plane of the torus. Decomposed into circular components,
such a field can be written in global coordinates as:
\begin{equation}
\mathbf{B}_{2}= \frac{u}{\sqrt{2}}\left(\begin{array}{c}
1\\
i\\
0\\
\end{array}\right)+\frac{v}{\sqrt{2}}\left(\begin{array}{c}
1\\
-i\\
0\\
\end{array}\right)
\end{equation}
Using Eq.(\ref{eqSigmaComponent}), the corresponding coupling
element in local coordinates is given by:
\begin{equation}
\beta_{+}=\frac{u}{2}\,(1+\sin\varphi)\, i\,e^{i\theta}
-\frac{v}{2}\,(1-\sin\varphi)\, i\,e^{-i\theta}
\end{equation}
Due to the direction of the static field
$\mathbf{B}_{\mathrm{dc}}$, these components do not couple at
either the top or the bottom of the torus, depending on their
sense of rotation. On the opposite side, they interfere with the
cylindrical field $\mathbf{B}_1$, leading to maxima and minima
along the toroidal angle $\theta$. As the direction of the
interference is opposite for $u$ and $v$, a linearly polarized
field ($|u|=|v|$) with a frequency different from that of
$\mathbf{B}_1$ will induce equal but counterpropagating rotations
of the two rings. Resulting potentials are shown in Fig.
\ref{figPotCounterRings}. It is interesting to mention that this
configuration can be used for state-dependent control of atomic
motion \cite{hofferberth}, because the rotational sense of the
coupling field depends on the sign of the Land\'e factor $g_F$.
Due to the symmetry, the field $\mathbf{B}_1$ creates a
state-independent potential for states with the same modulus
$|g_F|$, while the circular components of field $\mathbf{B}_2$
swap roles for opposite signs of $g_F$. We note that the field
$\mathbf{B}_2$ can also be used to cancel potential differences
due to a tilt in the gravitational field.

\section{Conclusion and outlook}

The possibility to split the torus into two rings with a
dynamically tunable barrier enables tunneling (Josephson
oscillations) and superconducting-quantum-interference-device-
\textsc{squid-} type experiments. Several rf frequencies
\cite{courteille} can be used to create independently controlled,
concentric atom shells, providing a tool for interferometric
analysis and for the creation of exotic lattices.

In conclusion, we have presented a class of rf dressed magnetic
traps that offer great flexibility and a high degree of dynamical
control. This versatility arises from the combination of a static
ring-shaped quadrupole field with rf-fields of suitable
polarizations. It enables studies of low-dimensional gases with
periodic boundary conditions, creation of vortex loops, persistent
currents, and solitons. Coherent splitting of matter-waves
\cite{schumm} together with independent control of atomic motion
in split rings, as well as state-dependent control of atomic
motion makes this trap type particularly interesting for rotation
sensing \cite{clauser}.

\section{Acknowledgements}

We gratefully acknowledge helpful discussions with Mikhail Baranov
and with the Heidelberg group. This work is part of the research
program of the Foundation for Fundamental Research on Matter
(FOM), which is financially supported by the Dutch Organization
for Scientific Research (NWO). It was also supported by the EU
under contract MRTN-CT-2003-505032 (Atom Chips). P.\,K.
acknowledges support by the Alexander von Humboldt-Stiftung and
the EU, under Contract No. MEIF-CT-2006-025047.


\end{document}